\documentclass[a4paper,11pt]{article}
\usepackage{pos}
\usepackage{slashed}
\usepackage{tikz-feynman}
\usetikzlibrary{external}
\immediate\write18{mkdir -p pgf-img}
\title{$\ell N$ inclusive scattering cross sections on the lattice}

\author*[a]{Jun-Sik Yoo}
\author[a,b]{Shoji Hashimoto}
\author[c,d]{Hiroshi Ohki}

\affiliation[a]{Theory Center, Institute of Particle and Nuclear Studies, High Energy Accelerator Research Organiztion (KEK),
  1-1 Oho, Tsukuba 305-0801, Japan }

\affiliation[b]{School of High Energy Accelerator Science, The Graduate University for Advanced Studies (SOKENDAI),\\
Tsukuba, Ibaraki 305-0801, Japan}
\affiliation[c]{Department of Physics, Nara Women's University,\\
Nara 630-8506, Japan}
\affiliation[d]{RIKEN/BNL Research Center, Brookhaven National Laboratory,\\
Upton, NY 11973, USA}

\emailAdd{junsik@post.kek.jp}

\abstract{Utilizing the approach recently proposed for the $\ell N$ inclusive scattering cross section on the lattice, we compute the differential scattering cross section for the charged current process $\ell p \rightarrow \nu n $ for various kinematical channels. The simulation is carried out on the 2+1-flavor $16^3 \times 32$ ensemble with Iwasaki and Domain Wall Fermion action. The lattice results are compared with MINER$\nu$A result for the equivalent process. }

\FullConference{%
 The 38th International Symposium on Lattice Field Theory, LATTICE2021
  26th-30th July, 2021
  Zoom/Gather@Massachusetts Institute of Technology
}

\begin{document}
\maketitle

\section{Introduction}
Accurate measurement of the neutrino cross section is important for many topics in physics such as reducing the background events in the rare decay measurement, or performing precision measurement of the neutrino oscillation parameters. Well-controlled neutrino sources have become available recently, and they are used to measure the most precise cross section for the neutrino-nucleon scattering.  Currently, the $\nu N$ scattering is actively investigated in projects such as T2K \cite{T2K:2019dgm}, MINER$\nu$A \cite{MINOS:2014axb,MINERvA:2018hqn,MINERvA:2019gsf}, and SciBooNE \cite{Szelc:2015tua,MicroBooNE:2019nio}. However, the fully non-perturbative theoretical computation of the total scattering cross section has never been done because it requires results from all ranges of energy, and each of the processes in the different energy regime gives distinct and significant challenges for the theoretical calculation of the scattering cross section for the process.

In the low energy regime, the quasi-elastic (QE) process is dominant for which one can use the form factor decomposition. In an intermediate energy scale, the pion and other particles may be generated possibly through resonance states, making it more complicated to analyze. In the high scale, it gradually approaches the deep inelastic scattering (DIS) process, where one needs the parton distribution function as well as the fragmentation function, but the factorization \cite{Collins:1989gx} of long- and short-range interaction is not obvious, especially for higher twist contributions which become relevant beyond the leading order of the operator product expansion (OPE). Such difficulties have been an impediment one has to overcome to give a unified and consistent view on the process.

This process can be decribed by lattice QCD, on the other hand, from the first principles of QCD in a consistent way over the whole regime \cite{Fukaya:2020wpp}. We propose to use the Chebyshev approximation of the kernel function, to perform the energy integral necessary to the sum over all possible final states in the inelastic scatterings.

\section{Inclusive scattering formalism}
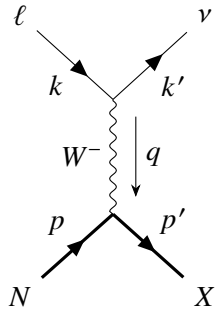
\begin{figure}[b]
  \centering
  \feynmandiagram [vertical'=a to b] {
    i1 [particle=\(\nu\)]
      -- [anti fermion, edge label=\(k^{\prime}\)] a
      -- [anti fermion, edge label=\(k\)] f1 [particle=\(\ell\)],
    a -- [boson,  momentum=\(q\), edge label'=\(W^{-}\)] b,
    i2 [particle=\(N\)]
      -- [fermion, edge label=\(p\), very thick] b
      -- [fermion, edge label=\(p^{\prime}\), very thick] f2 [particle=\(X\)]
  };
  \caption{Nucleon-lepton scattering }
  \label{fig:feyndiag}
\end{figure}

The formalism is outlined in Ref.~\cite{Gambino:2020crt,Fukaya:2020wpp}. 
For the $\ell N$-scattering process (Fig.~\ref{fig:feyndiag}),
the total cross section can be written as:
\begin{equation}
\sigma  = \int d^3 \mathbf{q} \int d \omega K(\omega ; \boldsymbol{q}) \rho(\omega ; \boldsymbol{q}),
\label{eq:master}
\end{equation}
while the lattice correlator is written as
\begin{equation}
C(t ; \boldsymbol{q})=\int_{0}^{\infty} d \omega \rho(\omega ; \boldsymbol{q}) e^{-\omega t},
\label{eq:correlator}
\end{equation}
with the spectral function 
\begin{equation}
\rho(\omega ; \boldsymbol{q}) \propto \sum_{X(\boldsymbol{q})} \delta\left(\omega-E_{X(\boldsymbol{q})}\right)|\langle X(\boldsymbol{q})|J| N\rangle|^{2},
\end{equation}
where $J$ is a local current to induce the scattering, $\hat{H}$ is the Hamiltonian operator, and $| N \rangle $ is a nucleon state. The sum runs over all possible states $X(\boldsymbol{q})$ with a specified momentum $\boldsymbol{q}$. We take $\tilde{J}(\mathbf{q}) \equiv \sum_{\boldsymbol{x}} e^{-i \boldsymbol{q} \cdot \boldsymbol{x}} J(\boldsymbol{x}) $, is a Fourier transformed interpolator which carries a spatial momentum of $\boldsymbol{q}$. 
Then we approximate the integral in Eq.~\eqref{eq:master},
\begin{align}
  \int d \omega K(\omega ; \boldsymbol{q}) \rho(\omega ; \boldsymbol{q}) \propto \langle N|\tilde{J}(-\boldsymbol{q}) K(\hat{H} ; \boldsymbol{q}) \tilde{J}(\boldsymbol{q})| N(\mathbf{0})\rangle,
\end{align}
using the polynomial of $e^{-\hat{H}}$,
\begin{equation}
K(\hat{H} ; \boldsymbol{q}) \simeq k_{0}(\boldsymbol{q})+k_{1}(\boldsymbol{q}) e^{-\hat{H}}+k_{2}(\boldsymbol{q}) e^{-2 \hat{H}}+\cdots+k_{N}(\boldsymbol{q}) e^{-N \hat{H}},
\end{equation}
which can be constructed using the correlator Eq.~\eqref{eq:correlator}.

\section{Contractions}
\begin{figure}[tbp]
  \centering
  \includegraphics[width=.33\textwidth]{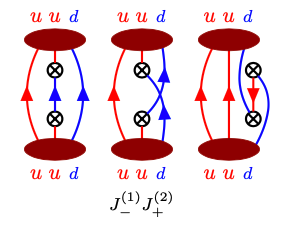}
  \caption{All allowed diagrams for $\langle N J^{-} J^{+} \bar{N} \rangle $. Red ellipse is the source and the sink nucleon, $\otimes$ mean the current insertion, and red and blue quark lines are the u and d quarks, respectively.}
  \label{fig:PM_diagram}
\end{figure}

In order to compute the forward Compton amplitude corresponding to the $\ell N$ scattering process, we need the Wick contraction for the two current insertion with nucleon initial and final states. For the charged current (CC) process $e p \rightarrow \nu n$, for example, the desired forward Compton scattering amplitude can be written as below or represented as in the diagrams in Fig.~\ref{fig:PM_diagram}:
\begin{equation} 
  \langle n | \tilde{J}^{-}_{\mu}(-\mathbf{q}) \tilde{J}^{+}_{\nu} (\mathbf{q}) | p \rangle.
\end{equation}
We have implemented the current insertion of every combination for the lepton-nucleon scattering, and any other channels can be analyzed with the same method. However, we focus on the CC process only in this work. 

\begin{figure}
  \centering
  \includegraphics[width=.4\textwidth]{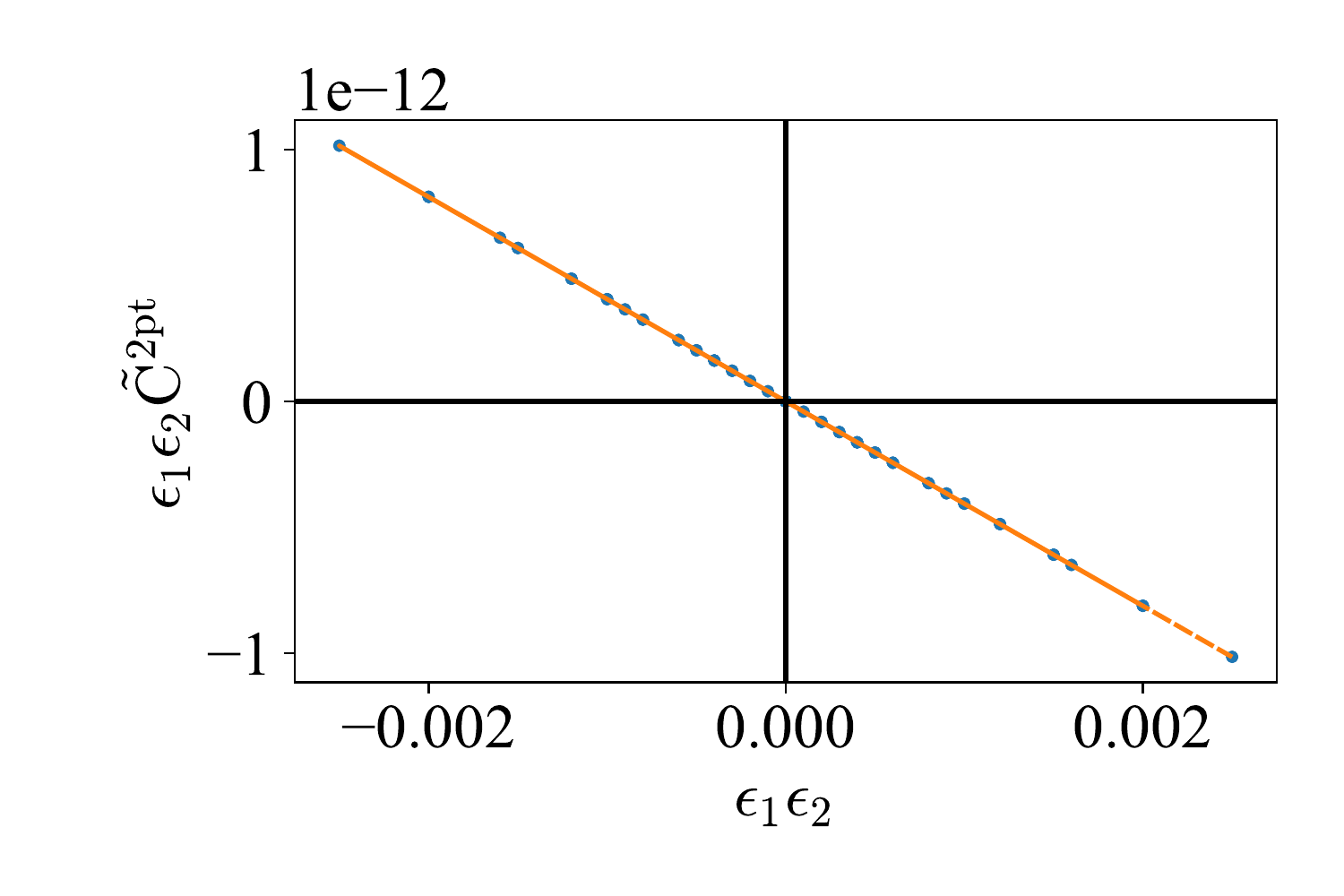}%
  \includegraphics[width=.4\textwidth]{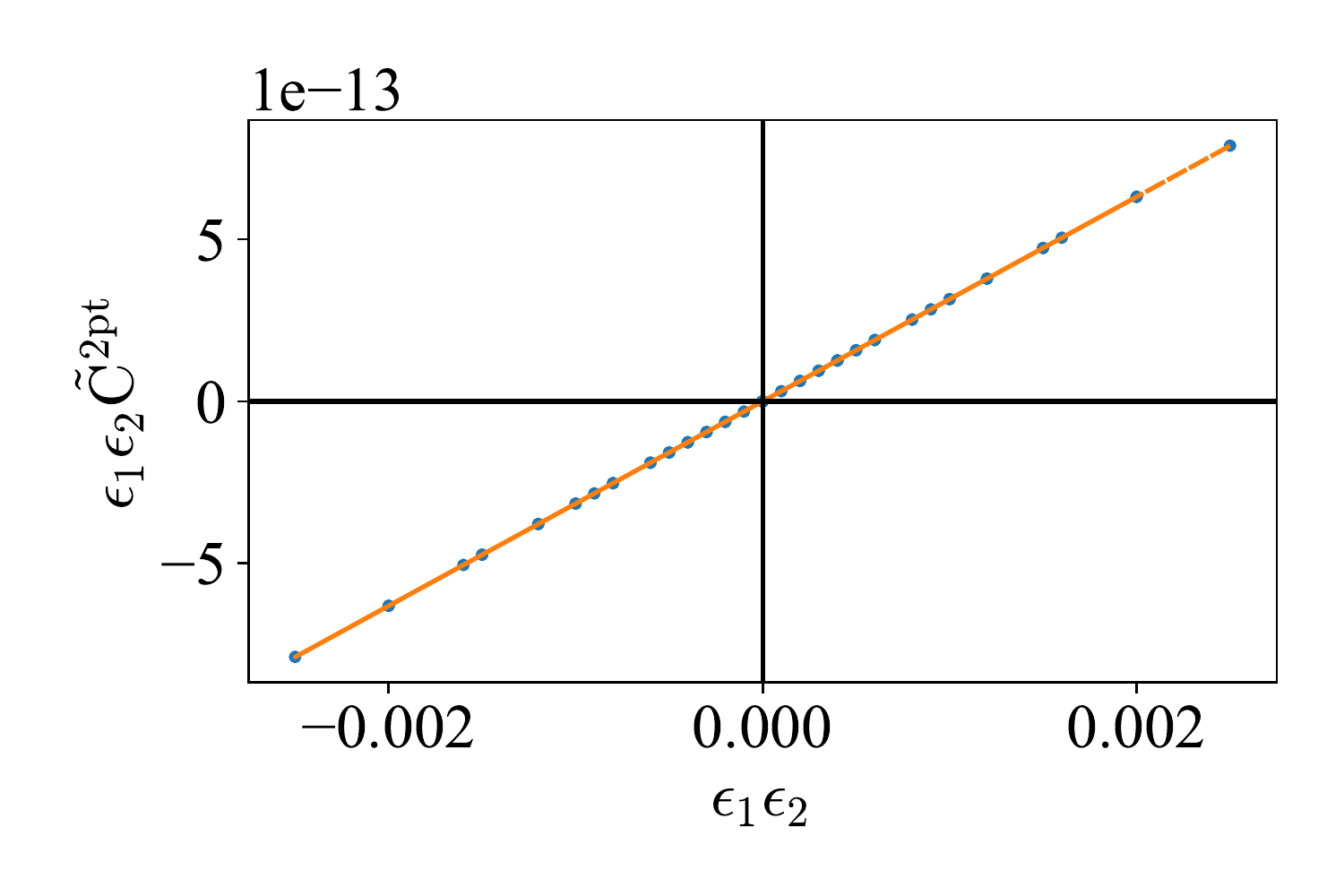}
  \caption{(left) Real part and (right) Imaginary part of $\langle P(t=8) \bar{P}(0) \rangle(\epsilon_1, \epsilon_2) $, where $\mathbf{q} = [0 0 1]$ in the lattice unit. }
  \label{fig:2pt_4pt}
\end{figure}

The contraction for the two-current insertion requires due test. We compare the four-point function results with the result from perturbed two-point functions.  Let us consider the current with an external field $\mathcal{A}_\mu$. The effective Lagrangian can be expressed:
\begin{equation}
\mathcal{L}_{\text {eff }}=\mathcal{L}_{\text{QCD}}+\epsilon_{1} J_{\mu}^{(1)} \mathcal{A}^{\mu}+\epsilon_{2} J_{\mu}^{(2)} \mathcal{A}^{\mu},
\end{equation}
where $\epsilon_{1,2}$ are some small parameters. On this background field, we replace the Dirac operator $\slashed{D}$ by 
\begin{equation}
\slashed{D} \rightarrow\left(\slashed{D}+\epsilon_{1} \Gamma^{(1)}+\epsilon_{2} \Gamma^{(2)}\right)
\end{equation}
The nucleon two-point function with given Lagrangian writes:
\begin{align}
  \langle N \bar{N} \rangle^{pert.}(\epsilon_1, \epsilon_2) 
  &=  \langle  N \bar{N} \rangle + \epsilon_1 \langle N J^{(1)} N \rangle 
  + \epsilon_2 \langle N J^{(2)} N \rangle + \epsilon_1 \epsilon_2 \langle N J^{(1)} J^{(2)} N \rangle + O(\epsilon_1^2, \epsilon_2^2)
\end{align}
Then, we take the derivative with respect to $\epsilon_{1,2}$ of the perturbed nucleon two point functions, so that we have the four-point functions. 
\begin{equation}
\lim_{\epsilon_1, \epsilon_2 \rightarrow 0 }\frac{\partial^2}{\partial \epsilon_1 \partial \epsilon_2} \langle N \bar{N} \rangle (\epsilon_1, \epsilon_2)
  =  \langle N J_{\mu}^{(1)}  J_{\nu}^{(2)} \bar{N} \rangle
\end{equation}
A more detailed explanation can be found in the Ref.~\cite{Fukaya:2020wpp}. One can make a linear combination of $C^{2pt}$'s to have only the relavant terms, $\epsilon_1 \epsilon_2 \tilde{C}^{2pt} $. We compute the  $C^{2pt}_{00}(t_{snk}=8, t_1=5, t_2=2, t_{src}=0 ; \mathbf{q}, \mathbf{p} ; \epsilon_1, \epsilon_2)$ in the perturbed two-point function method on the $16^3 \times 32$ lattice ensemble for comparison (Fig.~\ref{fig:2pt_4pt}), where $\mathbf{p}, \mathbf{q}$ were momenta insertion for the nucleon and the current to give the four-point function $\langle P(t=8) J^{+}_0(t_1 = 5 ; -\mathbf{q}) J^{-}_0(t_2 = 2 ; \mathbf{q} ) \bar{P}(0) \rangle $ after taking derivatives. The values of $\epsilon_{1,2}$ are given in between -0.05 and 0.05 with a separation of 0.01. 

The slope of the Fig.~\ref{fig:2pt_4pt} are $-4.062220\times 10^{-10} $ and $3.154343 \times 10^{-10}$ respectively,
whereas $-4.062221 \times 10^{-10}$ and  $3.154344\times 10^{-10} $ are the real and imaginary part of the four-point function values respectively.

\section{Simulation}
\subsection{Lattice setup}
We perform our QCD calculation on a RBC/UKQCD generated lattice ensemble \cite{RBC-UKQCD:2008mhs} of size $16^3 \times 32 $ using a heavier quark mass with chirally symmetric action. 
The $\beta$ and the lattice cutoff ($a^{-1}$) is 2.13 and 1.73(3) GeV \cite{RBC-UKQCD:2008mhs}, respectively. The pion mass is $399(10)$ MeV and $m_\pi L$ is 3.69, and the number of configurations are 139 and we take one exact sample per configuration. 
To reduce the cost of simulation we use the "zMobius" fermion action of $L_s = 10$. 
The renormalization constant for the axial current is $Z_A = 0.7162(2)$ \cite{RBC:2007yjf} and we use the same value for the $Z_V$ considering the error margin of our calculation.
For the source and the sink, interpolating operators are Gaussian smeared with $N= 70, \alpha= 0.5625$. 
In the simulation, the source-sink separation is fixed to 8, and the first current is inserted at $t=2$ slice. 

\subsection{Approximation of the kernel function}
While the polynomial term of $\omega$ consists of the kernel function, we implement the integration range to the integrand using the step function so that the integration range of $\omega$ is extended to the infinity. Here, we define the smeared kernel function as a product of polynomial term in $\omega$, smeared step function with a width $\sigma$, 
\begin{equation}
  \bar{K}^l(\omega) = e^{2\omega t_0} \omega^l \times \theta_{\sigma} (   m_N+ |\mathbf{q}| - \omega ).
  \label{eq:smoothed_kernel}
\end{equation}

\begin{figure}
  \vspace{-2cm}
  \hspace{-1.6cm} \includegraphics[width=1.22\textwidth]{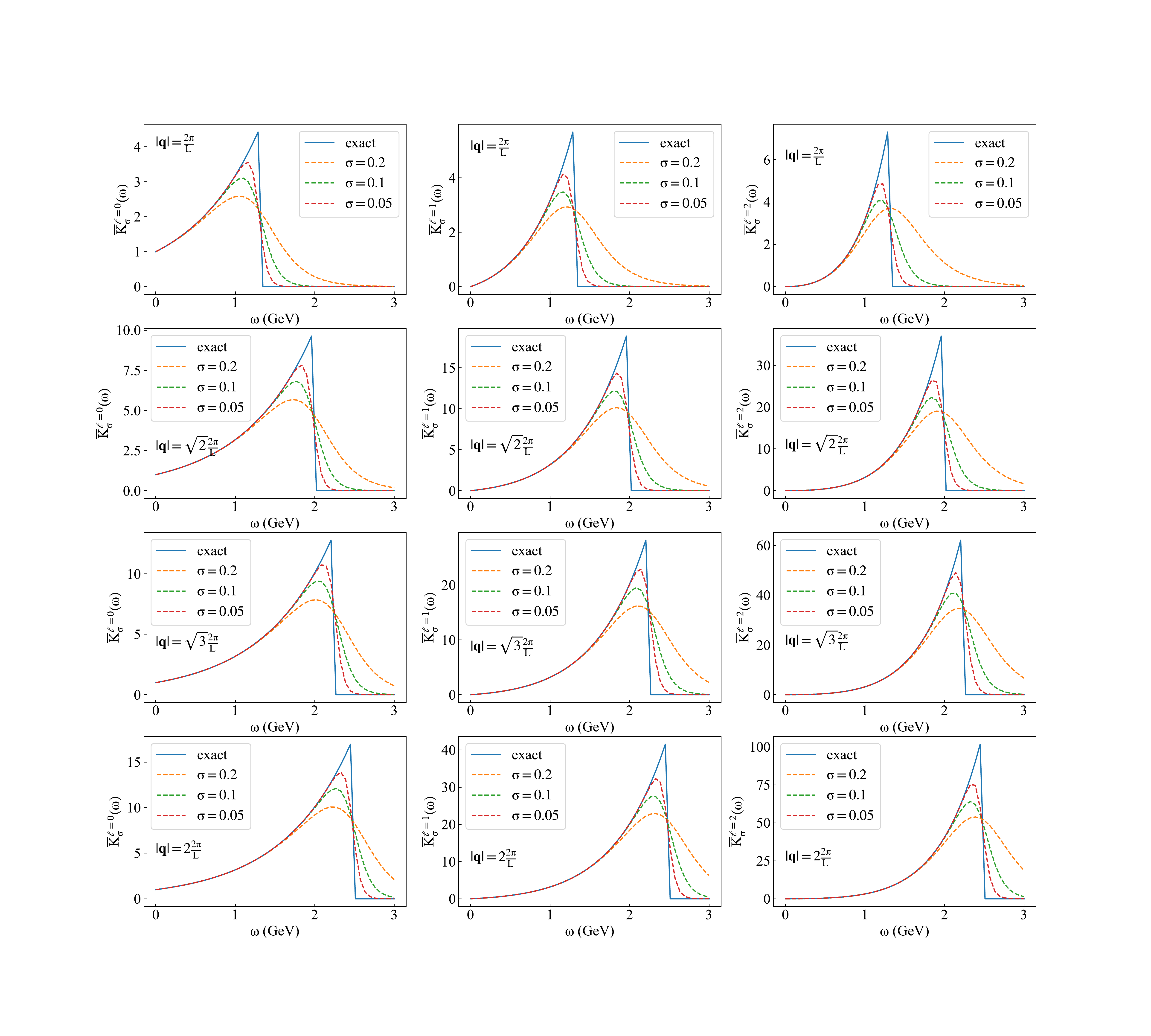}
  \vspace{-1.4cm}
  \caption{Smeared kernel functions of (Left column) $l = 0$, (Center column) $l =1 $, (Right column) $ l = 2$ with different insertion of momenta.
          Transfer momenta is $|\mathbf{q}| = \frac{2\pi}{L} \times \sqrt{m}$ with (First row)  $m=1$, (Second row) $m=2$, (Third row) $m=3$, (Fourth row) and $m=4$. }
  \label{fig:kernel_smearing}
\end{figure}

\begin{figure}
  \centering
  \includegraphics[width=\textwidth]{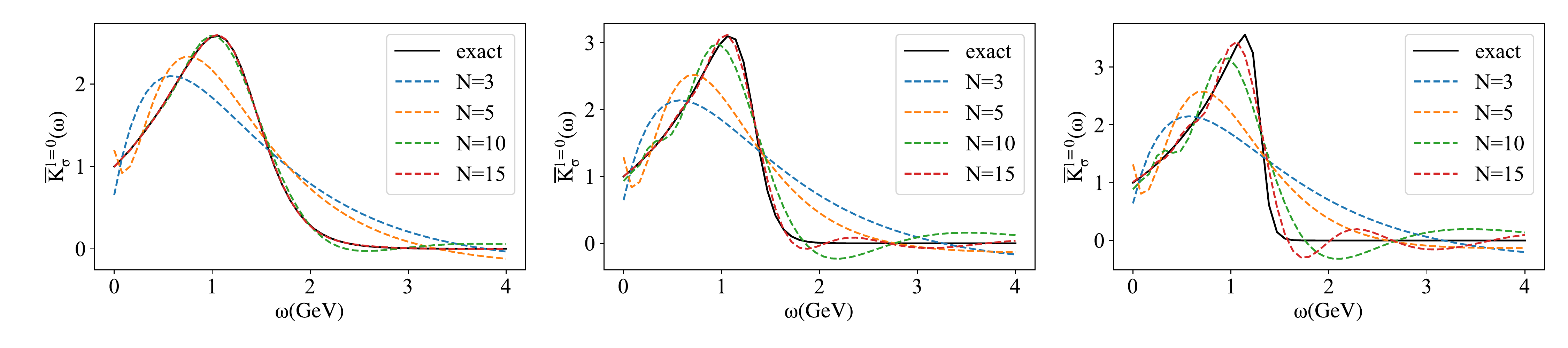}
  \caption{Chebyshev approximation of smeared kernel functions with (Left) $\sigma =0.2$ GeV (Center) $\sigma = 0.1$ GeV (Right) $\sigma = 0.05 $ GeV at different orders $N = \{3,5,10,15\}$. Narrow width requires higher order to be exactly approximated.}
  \label{fig:cheb_approx}
\end{figure}

The exact kernel function for the nucleon of mass $ 1.29 $ GeV at different orders of $\omega$ in the polynomial term ($l = 0, 1, 2) $ and with different momentum insertion $|\mathbf{q}| = \frac{2\pi}{L} \times \sqrt{m}, m=1,2,3,4$ are approximated with smeared kernel functions of different smearing widths $\sigma = 0.2$ GeV, $0.1$ GeV, and $ 0.05 $ GeV (Fig.~\ref{fig:kernel_smearing}).
For each of smeared kernel functions, we approximate with Chebyshev polynomials at different orders, $N=3,5,10,$ and $15$ (Fig.~\ref{fig:cheb_approx}). 

In the $\ell N$-scattering process, the prefactor and the leptonic tensor combined are the kernel function from the expression of the double differential cross sections,
\begin{equation}
  \frac{d^2 \sigma }{dQ^2 d\omega} = \frac{G_F^2 \omega }{4\pi E^2 m_N} L_{\mu \nu} W^{\mu \nu}.
\end{equation}
The kernel function is then $ K_{\mu \nu} = \frac{G_F^2 \omega }{4\pi E^2 m_N} L_{\mu \nu}$, where 
\begin{equation}
  L_{\mu \nu} = 2[ k'_\mu k_\nu + k_\mu k'_\nu - k' \cdot k g_{\mu \nu} + i \epsilon_{\mu \nu \alpha \beta } k^{\alpha } k'^{\beta}].
\end{equation}
Note that, the prefactor has additional term $ \frac{m_{W}^{4}}{\left[\left(\omega-m_{N}\right)^{2}-\mathbf{q}^{2}\right]^{2}}$ for the EM current case because the photon is massless unlike the $W, Z$ bosons. 

For the kernel function, we use
\begin{align}
  K^{\mu \nu}(\omega, \mathbf{q}) &=  \frac{G_F^2 \omega }{4\pi E^2 m_N} L_{\mu \nu} (E, \mathbf{q}) \nonumber \\
    &= C_{\mu \nu}^2 (\omega) \bar{K}^{ l = 2 }(\omega) \cdot \omega^2 + C_{\mu \nu}^1(\omega) \bar{K}^{ l = 1 }(\omega) \cdot \omega
     + C_{\mu \nu}^0 (\omega) \bar{K}^{ l = 0 }(\omega) .
     \label{eq:kernel_def}
\end{align}
We take the Chebyshev polynomials,
\begin{equation}
  \bar{K}^l (\omega ) \simeq \frac{c^*_0(\mathbf{q})}{2} + \sum_{j=1}^{N} c_j^*(\mathbf{q}) T^*_j (z)
\end{equation}
where $z = e^{-\omega}$ and  $T_j^*$'s are shifted Chebyshev polynomial.
The coefficient for the approximation of correlator are obtained by the formula
\begin{equation}
  c_j^*(\mathbf{q}) = \frac{2}{\pi} \int^{\pi}_{0} d \theta K\left( - \ln \frac{1+ \cos \theta }{2} ; \mathbf{q} \right) \cos(j\theta)
\end{equation}
Thus, employing correlator values 
\begin{equation}
  \frac{C_{\mu \nu}(t+2t_0)}{C_{\mu \nu}(2t_0)} = \frac{\langle \psi_{\mu} (\mathbf{q}) | e^{-\hat{H} t} | \psi_{\nu} (\mathbf{q}) \rangle }{\langle \psi_{\mu} (\mathbf{q}) | \psi_{\nu} (\mathbf{q}) \rangle} = \langle e^{-\hat{H}t} \rangle \sim \hat{z}^t
\end{equation}
at each order we make a basis consisting of Chebyshev polynomial $T^*_j(\hat{z})$. The maximum order of $\hat{z}$ is determined by the separation between two currents, which is $t_2 = \{4,5,6\}$ in our case.

Now, we consider all things together,

\begin{align}
  \int d\omega &\bar{K}^l(\omega)  \int \frac{d^3 \mathbf{P}_X}{(2\pi)^3}\sum_{X(\mathbf{P}_X)} \delta( \omega - E_{X(\mathbf{P}_X)}) 
  (2\pi)^3 \delta^{(3)}(\mathbf{P}_X - \mathbf{q}) \nonumber \\
  & \times \langle N | J^\mu(-\mathbf{q}) |  X(P_X) \rangle \langle X(P_X) | J^{\nu}(\mathbf{q}) | N \rangle \\
  &= \langle N |  J^\mu_X(-\mathbf{q})  \bar{K}^l ( \hat{H}) J_Y^{\nu}(\mathbf{q}) | N \rangle \equiv I^{\mu \nu}_{XY, l} (\mathbf{q}),
  \label{eq:single_kernel}
\end{align}
where $X, Y = \{ V, A \}$.

\section{Result}

\begin{figure}
  \centering
  \includegraphics[width=.91\textwidth]{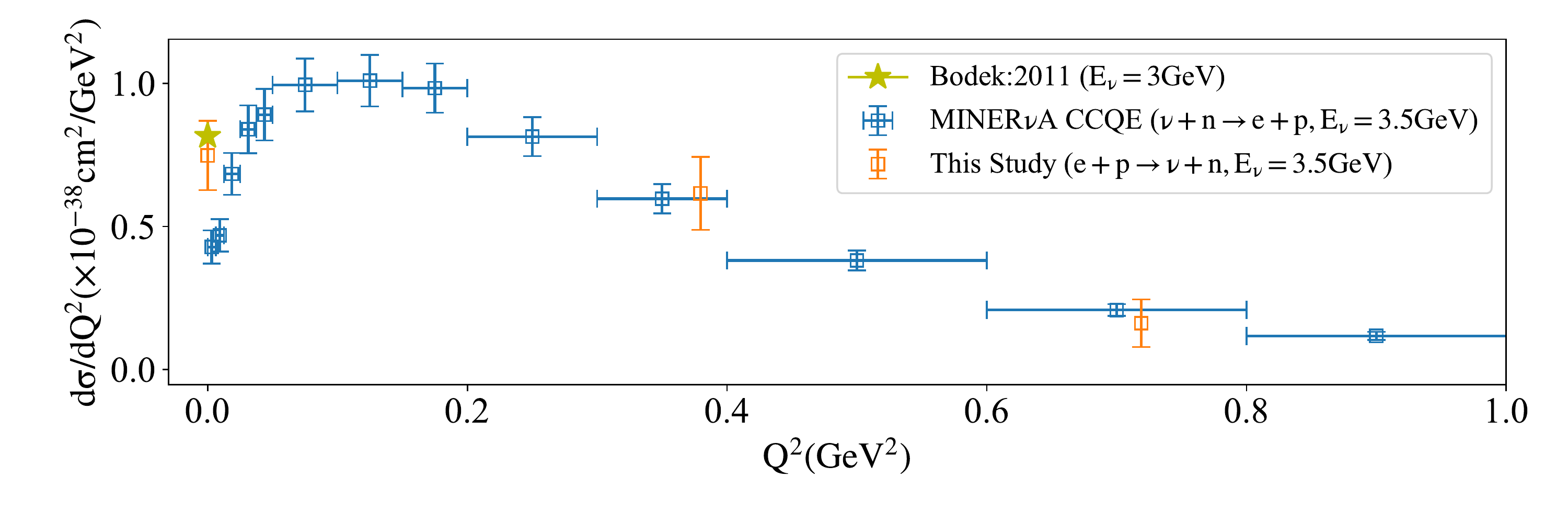}
  \caption{$Q^2$-differential cross section is investigated in model calculation (Bodek:2011 \cite{Bodek:2011ps}), experimental result (MINERvA \cite{MINERvA:2018hqn}), and the lattice calculation (This study). }
  \label{fig:q2_sc}
\end{figure}

We first compute all the $I^{\mu \nu}_{XY, l}$ values and take the linear combination. By combining Eq.~\eqref{eq:single_kernel} and Eq.~\eqref{eq:kernel_def},
\begin{align}
  \frac{d\sigma(E)}{dQ^2} &=   \sum_{\mu, \nu, l} C^{l}_{\mu \nu} I^{\mu \nu}_{XY, l} (\mathbf{q}).
\end{align}
The differential scattering cross section for the CC process $ \ell p \rightarrow \nu n $ is shown in the Fig.~\ref{fig:q2_sc} as a function of $Q^2$, together with the experimental data of the quasi-elastic cross sections \cite{MINERvA:2018hqn,Bodek:2011ps}. We set the energy of the incoming neutrino as $E_\nu = 3.5$ GeV.

Although the result in Fig.~\ref{fig:q2_sc} looks reasonable, there are a few caveats about the computation. First, the order of the Chebyshev approximation is only $N=3$, because of the temporal size of the lattice. The order of the Chebyshev approximation depends on the source-sink separation, and it is limited by the temporal lattice size. The truncation at $N=3$ in our simulation introduces $O(10^{-2})$ systematic errors when the uniform spectral function over the energy range of $ [0\text{ GeV}, 4\text{ GeV}] $ is assumed. However, we do not have contribution from below the kinematic cut and the effect of shifted peaks from the truncation is not well incorporated in this number. To investigate the systematic errors in the approximation, a more detailed approach is needed.

The volume of the lattice ensemble is not enough to reproduce the bump in the differential cross section in the low energy region. Larger spatial size of the lattice will allow finer configurations of transfer momenta. 

Another caveat is that the heavier nucleon of mass $1.29 $ GeV is used in the simulation, and our result is subject to the chiral extrapolation and its systematic errors. 

\section{Discussion}

This is the first application of the methodology to compute the inelastic $\ell N$ scattering process \cite{Fukaya:2020wpp}. A natural extension of this study is to perform calculations on larger volumes. It is crucial to have enough separations between the two currents in order to better control the Chebyshev approximation of the energy integral. Another extension is to approach the high energy regime by setting the lepton energy higher. In principle, the same lattice result can be applied to analyze the high energy process.  Since the methodology deals with the energy integral, the 1-loop correction to the nucleon $\beta $ decay process, namely the $W \gamma $ exchange contribution is also accessible with the same set of the data, when it is organized differently in the Wick contraction level. In order to truly evaluate the whole energy regime, employing the multigrid method with fine and coarse lattices would be necessary. 

\acknowledgments We thank the members of the JLQCD collaboration for helpful discussions and providing the
computational framework and lattice data. 
The computations were performed using the Qlua software suite \cite{qlua} with (z)M\"obius solvers from the Grid library \cite{grid}. The gauge configurations
have been generously provided by the RBC/UKQCD collaboration. Numerical calculations are performed on SX Aurora
Tsubasa at High Energy Accelerator Research Organization (KEK) under its Particle Nuclear and
Astrophysics simulation program as well as on Oakforest PACS supercomputer operated by Joint
Center for Advanced High Performance Computing (JCAHPC). This work is supported in part by JSPS KAKENHI Grant Numbers 17K14309, 18H03710, and 21K03554.

\bibliographystyle{JHEP}
\bibliography{ref}

\end{document}